\documentclass[
aps,
 ,twocolumn
]{revtex4}
\usepackage{graphicx}
\usepackage{amsmath}

\begin{document}

\title{Spacetime diamonds}

\author{Daiqin Su}\email{d.su@uq.edu.au}
\author{T.~C.~Ralph}\email{ralph@physics.uq.edu.au}
\affiliation{Centre for Quantum Computation and Communication Technology, \\
School of Mathematics and Physics, University
of Queensland, Brisbane, Queensland 4072, Australia}

\pacs{03.70.+k, 03.65.Ud, 04.62.+v}

\date{\today}


\begin{abstract}
{We show that the particle-number distribution of diamond modes, modes that are localized in a finite space-time region, are thermal for the 
Minkowski vacuum state of a massless scalar field, an analogue to the Unruh effect.  
The temperature of the diamond is inversely proportional to its size. An inertial observer can detect this thermal radiation by 
coupling to the diamond modes using an appropriate energy-scaled detector. 
We further investigate the correlations between various diamonds and find that
entanglement between adjacent diamonds dominates. }
\end{abstract}

\maketitle

\vspace{10 mm}

\section{Introduction}
A key result of relativistic quantum field theory is that the restriction of observers to partial regions of spacetime leads to the observation 
of particles, even if the total spacetime is in the vacuum state (see Ref. \cite{Peres04} and references therein). Key examples are Hawking radiation \cite{Hawking75}, where
the observers are cut off from the inside of a black hole by its event horizon, and Unruh-Davies radiation \cite{Unruh76, Davies75,Crispino08},
where uniform acceleration of the observer restricts them to a Rindler wedge through the formation of a virtual horizon. 
Both Hawking and Unruh-Davies radiation are thermal and their
temperatures are proportional to the surface gravity of the black hole and the acceleration of a uniformly accelerated observer,
respectively. The thermal character of the radiation is closely related to entanglement of the observed field modes with others
hidden behind the horizon \cite{,Birrell84}. More recently, it has been predicted that particles should also be observed when a detector
is restricted to the future or past light cone \cite{Olson11}. In all these cases, the region the observer is restricted to is unbounded.
A natural question is whether an observer restricted to a bound region of spacetime can see thermal radiation. 
Using the thermal time hypothesis \cite{Rovelli94}, Martinetti and Rovelli 
found that an accelerated observer with a finite lifetime can experience an effective temperature, a generalization to the  Unruh-Davies
temperature \cite{Rovelli03}.  For the special case of an inertial observer
with a finite lifetime, the temperature at the middle of their lifetime is nonzero. This so-called ``diamond temperature" is given by \cite{Rovelli03}
\begin{equation}\label{diamond temperature}
T_D = \frac{2}{\pi \mathcal{T}} ,
\end{equation}
where $\mathcal{T}$ is the lifetime of the inertial observer. The diamond temperature 
arises because an observer with a finite lifetime does not have access to all the degrees of freedom of the quantum 
field. However, the temperature discovered by Martinetti and Rovelli is time dependent. Also, it was unclear what 
type of physical system could observe the diamond temperature. 

In Ref. \cite{Olson11}, an Unruh-DeWitt detector \cite{Unruh76, DeWitt80} with an energy scaling that effectively restricts it to the future or past light cone was shown to 
register a  thermal response identical to that of a uniformly 
accelerated Unruh-DeWitt detector. In this paper, we show that an Unruh-DeWitt detector with an energy scaling which effectively gives it a finite lifetime, or 
equivalently, confines it within one diamond, also registers a thermal response. The temperature that the detector sees 
is exactly the diamond temperature (\ref{diamond temperature}) discovered by Martinetti and Rovelli. We thus find a physical 
meaning for the diamond temperature: it is the temperature observed by a particular type of energy-scaled detector. 

This paper is organized as follows: we first introduce a new coordinate system to describe the spacetime 
inside a diamond in Sec. \ref{diamond coordinates}. 
In Sec. \ref{thermal radiation}, we derive the Bogoliubov transformation between the diamond modes and the Minkowski
modes, and calculate the particle-number distribution of the diamond modes in the Minkowski vacuum state. This is shown to correspond to
the diamond temperature. In Sec. \ref{detector response}, we explicitly show that an inertial Unruh-DeWitt detector with an energy scaling 
detects thermal radiation at the diamond temperature.
We discuss the entanglement between various diamonds in Sec. \ref{correlations} and conclude in Sec. \ref{conclusions}. 
In this paper,  we take $c=\hbar=k_B=1$ and the signature 
of Minkowksi metric as $(+,-,-,-)$.

\section{Diamond coordinates}\label{diamond coordinates}

A static observer with a finite lifetime stays at ${\bf r}=0$. The overlap of the future light cone 
of their birth and the past light cone of their death is called a diamond, satisfying $|t|+|{\bf r}| < 2/a$, where $2/a$ is the size of the 
diamond or $\mathcal{T}=4/a$ is the lifetime of the static observer. There exists a conformal transformation which maps
the diamond (bounded) to a Rindler wedge (unbounded) \cite{Rovelli03}. 
This motivates us to introduce a new coordinate system $(\eta, \xi, \zeta, \rho)$, called diamond coordinates, 
to describe spacetime events and field modes inside the diamond. The relationship between the diamond coordinates and Minkowski
coordinates is given by
\begin{equation}\label{CT}
\begin{split}
&\eta = \frac{1}{a}\tanh^{-1}\bigg\{\frac{a t}{1+a^2t^2/4-a^2r^2/4}\bigg\}, \\
&\xi = \frac{1}{a}\ln\bigg\{\frac{\sqrt{(1+a^2t^2/4-a^2r^2/4)^2-a^2t^2} }{f(t,x,y,z;a)}\bigg\}, \\
&\zeta = \frac{2y}{f(t,x,y,z;a)},  \\
&\rho = \frac{2z}{f(t,x,y,z;a)},
\end{split}
\end{equation}
where $f(t,x,y,z;a) = 1 - (at/2)^2 + (a r/2)^2 - a x$, and $r=\sqrt{x^2+y^2+z^2}$. Inside the diamond, the line element can be
written as 
\begin{equation}\label{metric}
ds^2 = \frac{4(d\eta^2 - d \xi^2) - e^{-2 a \xi}(d\zeta^2+d\rho^2)}{[\cosh(a\eta)+\cosh(a\xi)+\frac{a^2}{2}e^{-a\xi}(\zeta^2+\rho^2)]^2}.
\end{equation}
Although the $x$ direction appears special in the 
coordinate transformation [Eq. (\ref{CT})], no direction is preferred due to the rotational invariance of the diamond. 
In fact, the same conformal transformation \cite{Rovelli03} maps the region outside the diamond to another Rindler wedge. This nice property can help
us to intuitively understand correlations between field modes inside and outside the diamond. 

It can be shown that $ \zeta=\rho=0, ~\xi=const.$ are worldlines of uniformly accelerated observers 
with acceleration $\frac{a}{2}|\sinh(a\xi)|$ in the perspective of inertial observers. 
The most interesting one is $\zeta = \rho = \xi =0$, which is exactly the worldline 
of the static observer. Along the static worldline, $t = \frac{2}{a} \text{tanh}(\frac{1}{2}a \eta)$, 
or $d t = d\eta/\cosh^2(a\eta/2)$. That means the diamond clock ticks at the same rate as the inertial clock 
at $\eta = 0$, while the former ticks much faster than the latter when
 $\eta \rightarrow \pm \infty$.

\begin{figure}[ht!]
\centering
\includegraphics[width=8.0cm]{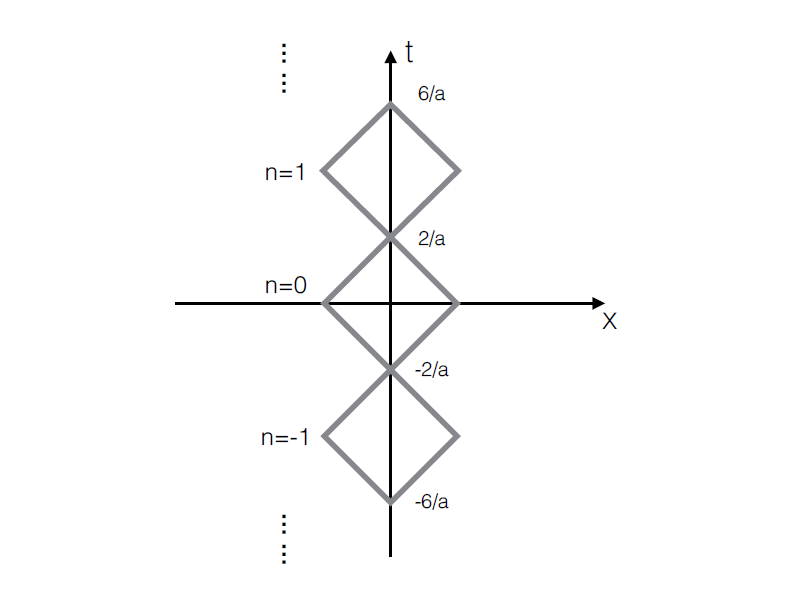}
\caption{\footnotesize Diamonds in $(1+1)$-dimensional Minkowski spacetime. Only a chain of diamonds along the 
$t$ axis is plotted and they are labeled by integers $n = 0, \pm 1, ... $.  The size of these diamonds are the same, $2/a$. } 
\label{D:Fig}
\end{figure}

\section{Thermal radiation}\label{thermal radiation}

As a concrete example, we first consider a massless Hermitian scalar field $\hat{\Phi}$ in the $(1+1)$-dimensional Minkowski
spacetime and directly calculate the Bogoliubov transformation between the diamonds modes and Minkowski plane wave modes.
We show that the Minkowski vacuum looks like a thermal state in the diamond and the temperature of the 
thermal state is inversely proportional to the lifetime of the static observer. 

A chain of diamonds along the $t$ axis is shown in Fig. \ref{D:Fig}. Other diamonds are not plotted, but one can imagine that
the $(1+1)$-dimensional Minkowski spacetime is in fact a network of such diamonds. Without loss of generality, we first consider
the zeroth diamond in Fig. \ref{D:Fig}. By simply setting $\zeta=\rho=0$ in Eq. (\ref{metric}), we can directly read out the metric
inside the diamond in terms of $\eta$ and $\xi$, which turns out to be conformal to the Minkowski metric (note that this is only 
true for $(1+1)$-dimensional spacetime). It is thus very easy to derive the Klein-Gordon
equation by utilizing the conformal invariance of the massless scalar field in $(1+1)$-dimensional Minkowski spacetime. 

Since the left-moving modes and right-moving modes are decoupled, we only discuss the left-moving modes in the following.
Results for the right-moving modes can be obtained similarly. The Minkowski annihilation operators and positive frequency 
mode functions are $\hat{a}_k$ and $u_{k}(V) = e^{- i kV}/\sqrt{4 \pi k}$, with $V=t+x$.
 While in the zeroth diamond they
are $\hat{b}^{(0)}_{\omega} $ and $g^{(0)}_{\omega}(v) = e^{- i \omega v}/\sqrt{4 \pi \omega}$, with $v=\eta+\xi$.
Meanwhile, $g^{(0)}_{\omega}(v)$ can be rewritten in terms of Minkowski null coordinate V,
\begin{equation} \label{modefunction}
g^{(0)}_{\omega}(V) = \frac{1}{\sqrt{4 \pi \omega}}\bigg( \frac{1+aV/2}{1-aV/2}\bigg)^{-i\omega/a},
\end{equation}
where $V \in [-2/a, 2/a]$ and the mode functions vanish outside the zeroth diamond.

We now have two ways to quantize the scalar field and it is straightforward to find the Bogoliubov transformation between operators
$(\hat{b}^{(0)}_{\omega}, \hat{b}^{(0)\dagger}_{\omega})$ and $(\hat{a}_{k}, \hat{a}^{\dagger}_{k})$,
\begin{equation}\label{Bogoliubov}
\hat{b}^{(0)}_{\omega} = \int d k \bigg( A^{(0)}_{\omega k} \hat{a}_k + B^{(0)}_{\omega k} \hat{a}^{\dagger}_k \bigg).
\end{equation}
Direct calculation shows that $B^{(0)}_{\omega k} \neq 0$, which means these two ways of quantization are inequivalent; in particular,
the Minkowski vacuum is not a vacuum in the diamond and vice versa. 
The Bogoliubov transformation coefficients
$A^{(0)}_{\omega k}$ and $B^{(0)}_{\omega k}$ can be calculated using the Klein-Gordon inner product \cite{Birrell84}; we have
\begin{equation}\label{Bogoliubov}
\begin{split}
&A^{(0)}_{\omega k} = \frac{1}{a}\frac{\sqrt{\Omega \kappa}}{\text{sinh}(\pi \Omega)} e^{2i\kappa}M(1+i \Omega, 2, -4i \kappa), \\
&B^{(0)}_{\omega k} = - \frac{1}{a}\frac{\sqrt{\Omega \kappa}}{\sinh(\pi \Omega)} e^{-2i\kappa}M(1+i \Omega, 2, 4i \kappa), 
\end{split}
\end{equation}
where $M(a,b,z)$ is the Kummer's function \cite{Abramowitz} and $\Omega \equiv \omega/a$, $\kappa \equiv k/a$. 
In the Minkowski vacuum state, the particle-number distribution in the diamond is
\begin{equation}\label{thermal distribution}
\langle 0 | \hat{b}^{(0)\dagger}_{\omega} \hat{b}^{(0)}_{\omega'} | 0 \rangle = \int d k B^{(0)*}_{\omega k} B^{(0)}_{\omega' k}  
= \frac{\delta(\omega - \omega')}{e^{2\pi \omega/a} - 1} ,
\end{equation}
which is exactly a thermal distribution with temperature
\begin{equation}\label{Rovelli}
T_D = \frac{a}{2\pi} = \frac{2}{\pi \mathcal{T}}.
\end{equation}
The temperature $T_D$ derived here is the same as the diamond temperature derived from the thermal time hypothesis \cite{Rovelli03}. 
The same thermal particle-number distribution was obtained by Ida {\it et al.} \cite{Ida13} through a different way. However, they use it as an intermediate 
result to derive the time-dependent temperature as proposed by Martinetti and Rovelli instead of 
interpreting it as the diamond temperature. 
We emphasize that $T_D$ is exactly the diamond temperature and will show that this thermal radiation could be detected by an 
energy-scaled Unruh-DeWitt detector. 

In principle, the above result can be generalized to $(1+3)$-dimensional spacetime. 
One can define diamond modes, calculate the Bogoliubov transformation coefficients and show that  the particle-number
distribution is thermal in the Minkowski vacuum. However instead of doing the long mathematical calculation, 
we propose a detector model in $(1+3)$-dimensional spacetime and show that it responds to the 
diamond temperature, which is more physically relevant.

\section{Detector response}\label{detector response}

We now turn to $(1+3)$-dimensional Minkowski spacetime. 
In Ref. \cite{Olson11}, an inertial detector switched on at $t=0$ and sensitive to energy $E$ with respect to conformal time 
is proved to register a thermal response. We now show that a similar inertial detector, which is switched on at $t = - \frac{2}{a}$ and
switched off at $t = \frac{2}{a}$, detects thermal radiation with diamond temperature
in the Minkowski vacuum. Because we require the energy difference of the two-level detector at ${\bf r}=0$ to be constant with respect to diamond time
$\eta$, the free Hamiltonian of the detector in the inertial frame should be $H_0/(1-a^2 t^2/4)$. We then take the Hamiltonian to be
$H = H_0/(1-a^2 t^2/4) + H_I$, where $H_I$ is the standard interaction term for an Unruh-DeWitt detector, $\lambda \hat{m} \hat{\Phi}$. 
Converting to diamond time $\eta$, the Schr\"{o}dinger equation is 
\begin{equation}
i \frac{\partial \Psi} {\partial \eta}= \bigg[H_0 + \frac{1}{\text{cosh}^2(a \eta/2)}H_I \bigg]  \Psi,
\end{equation}
where $\Psi$ is the wave function of the detector. In contrast to Ref. \cite{Olson11} where perturbation theory breaks down at sufficiently late time,
, the perturbation theory is always valid here provided $|H_I| \ll |H_0|$ at $\eta = 0$.
 
 To first order, the detector response function can be obtained in a standard way \cite{Birrell84}:
 \begin{equation}\label{response}
 \mathcal{F}(E) = \int_{-\infty}^{\infty} d \eta \int_{-\infty}^{\infty} d \eta'   \frac{e^{-i E(\eta - \eta')}D^+(\eta, \eta')}{\text{cosh}^2(a \eta/2)\text{cosh}^2(a \eta'/2)},
 \end{equation}
 where $D^+(\eta, \eta') = \langle 0 | \hat{\Phi}(\eta) \hat{\Phi}(\eta') | 0 \rangle$ is the positive-frequency Wightman function in the Minkowski 
 vacuum state. In terms of Minkowski coordinates $t$ and ${\bf r}$, the general expression of the Wightman function is \cite{Birrell84}
 \begin{equation}
 D^+(t,{\bf r}; t', {\bf r}^{\prime}) = -\frac{1}{4 \pi^2}\frac{1}{[(t-t'- i\epsilon)^2-({\bf r}-{\bf r}^{\prime})^2] }. 
 \end{equation}
 Taking into account that the inertial detector is at ${\bf r}=0$, and the relation between the Minkowski time and diamond time is
 $t = \frac{2}{a} \text{tanh}(\frac{1}{2}a \eta)$, we find 
 \begin{equation}\label{response-conformal}
 \frac{D^+(\eta, \eta')}{\text{cosh}^2(a \eta/2)\text{cosh}^2(a \eta'/2)} 
 = -\frac{1}{16 \pi^2} \frac{a^2 }{\text{sinh}^2(\frac{a}{2}(\eta-\eta'))}. 
 \end{equation}
 Now, consider an accelerated trajectory $t = a^{-1} \sinh(a \tau),~x = a^{-1} \cosh(a \tau),~y=z=0$, with $a$ and $\tau$ the proper 
 acceleration and proper time of the accelerated observer; in this case, the Wightman function is \cite{Birrell84}
 \begin{equation}\label{response-accelerated}
  D^+(t,{\bf r}; t', {\bf r}^{\prime})  = -\frac{1}{16 \pi^2} \frac{a^2}{\text{sinh}^2(\frac{a}{2}(\tau-\tau'))}. 
   \end{equation}
 Comparing Eqs. (\ref{response-conformal}) and (\ref{response-accelerated}), it is clear that the response function $\mathcal{F}(E)$ is 
 the same as that of a uniformly accelerated detector, 
 showing that an inertial Unruh-DeWitt detector with energy scaled as $\frac{1}{1-a^2 t^2/4}$ detects thermal radiation with 
 temperature $T_D = \frac{a}{2 \pi}$ in the Minkowski vacuum. 
 
 Energy-scaled detectors are physically realizable; e.g., by applying a time-dependent external electric field or magnetic field to an atom one 
 can realize a time-dependent Stark effect or Zeeman splitting. However, an order-of-magnitude estimate shows that for current technology the change of the electric 
 field or the magnetic field is not large and fast enough to detect the diamond temperature. More promising candidates might be artificial atoms
 such as the superconducting qubits and quantum dots \cite{Ralph15}.

\section{Correlations}\label{correlations}

In Minkowski vacuum state, a mode localized in the right Rindler wedge is perfectly entangled with a corresponding mode in the left
Rindler wedge \cite{Birrell84}. Similarly, a mode localized in the past light cone is perfectly entangled with a corresponding mode in the future 
light cone \cite{Olson11}. As we have mentioned before, there exists a conformal transformation that maps a diamond into a 
Rindler wedge, and the region outside the diamond to another Rindler wedge \cite{Rovelli03}. 
Therefore, if the dynamics of the scalar field is 
conformally invariant and also the vacuum state, then a mode inside the diamond should be perfectly entangled with a corresponding mode
outside the diamond. A pair of entangled modes inside and outside the diamond has been calculated in the $(1+1)$-dimensional 
spacetime \cite{Ida13}. Here, we are interested in the timelike entanglement between diamonds along the $t$ axis, which is now not
bipartite entanglement but multipartite entanglement. In this case, it is convenient to consider localized modes and introduce 
Gaussian formalism \cite{Weedbrook12} to describe the entanglement between various diamonds.



As shown in Fig. \ref{D:Fig}, orthonormal mode functions
in the $n$th diamond can be easily obtained by shifting those of the zeroth diamond, 
\begin{equation} 
g^{(n)}_{\omega}(V) = g^{(0)}_{\omega}(V-4n/a),
\end{equation}
where $V\in [2(2n-1)/a, 2(2n+1)/a]$. The Bogoliubov transformation coefficients $A^{(n)}_{\omega k}$ and $B^{(n)}_{\omega k}$ are
\begin{equation}
A^{(n)}_{\omega k} 
= e^{-4i n (\frac{k}{a})}A^{(0)}_{\omega k}, \,\,\,\,\,\,
B^{(n)}_{\omega k} 
= e^{4i n (\frac{k}{a})}B^{(0)}_{\omega k}. 
\end{equation}
It is obvious that the temperature in every diamond is the same, owing to the translational invariance of the Minkowski 
vacuum. 

Notice that the modes $g^{(n)}_{\omega}(V)$ are orthonormal and form a complete set of modes with which the scalar field 
$\hat{\Phi}$ can be expanded. The corresponding annihilation and creation operators are $\hat{b}^{(n)}_{\omega}$ and $\hat{b}^{(n)\dagger}_{\omega}$.
Another orthonormal and complete set of modes was introduced in Ref. \cite{Ida13}, the modes inside the zeroth diamond, $g^{(0)}_{\omega}(V)$, and 
that outside,
\begin{equation}
g^{(ex)}_{\omega}(V) = \frac{1}{\sqrt{4 \pi \omega}}\bigg(\frac{aV/2 + 1}{aV/2 - 1}\bigg)^{i\omega/a}\Theta(|V|-2/a),
\end{equation}
which is perfectly correlated with $g^{(0)}_{\omega}(V)$. By using this 
property and the Bogoliubov transformation between $g^{(ex)}_{\omega}(V)$ and $g^{(n)}_{\omega}(V)$ with $n \neq 0$, we can easily find
\begin{equation}
\begin{split}
&\langle 0 | \hat{b}^{(0)}_{\omega} \hat{b}^{(n)}_{\omega'} | 0 \rangle = \langle 0 | \hat{b}^{(0)\dagger}_{\omega} \hat{b}^{(n)\dagger}_{\omega'} | 0 \rangle^*
= \frac{\alpha^{(n)}_{\omega' \omega}}{2 \sinh(\pi \Omega)}, \\
&\langle 0 | \hat{b}^{(0)\dagger}_{\omega} \hat{b}^{(n)}_{\omega'} | 0 \rangle = \langle 0 | \hat{b}^{(0)}_{\omega} \hat{b}^{(n)\dagger}_{\omega'} | 0 \rangle^*
= \frac{\beta^{(n)}_{\omega' \omega}}{2 \sinh(\pi \Omega)}, 
\end{split}
\end{equation}
where $\alpha^{(n)}_{\omega' \omega} = \langle g^{(n)}_{\omega'}, g^{(ex)}_{\omega} \rangle$ and 
$\beta^{(n)}_{\omega' \omega} = \langle g^{(n)}_{\omega'}, g^{(ex)*}_{\omega} \rangle$, and $\langle \cdot, \cdot \rangle$ represents the 
Klein-Gordon product \cite{Birrell84}. 
For the $n=1$ case (adjacent diamonds), the coefficients $\alpha^{(1)}_{\omega' \omega}$ and $\beta^{(1)}_{\omega' \omega} $ can be calculated analytically:
\begin{equation}
\begin{split}
&\alpha^{(1)}_{\omega' \omega}  =  \frac{2^{i(\Omega-\Omega')}}{2 \pi a} \sqrt{\frac{\Omega}{\Omega'}}
\frac{\Gamma(1-i\Omega')\Gamma(i(\Omega'-\Omega))}{\Gamma(1-i\Omega)}, \\
&\beta^{(1)}_{\omega' \omega} = -\frac{2^{-i(\Omega+\Omega')}}{2 \pi a} \sqrt{\frac{\Omega}{\Omega'}}
\frac{\Gamma(1-i\Omega')\Gamma(i(\Omega'+\Omega))}{\Gamma(1+i\Omega)}.
\end{split}
\end{equation}
When  $\omega = \omega'$, $\alpha^{(1)}_{\omega' \omega}$ is divergent, which means the correlation between 
same-frequency modes is dominant. The divergence causes no problem, because it should be understood in the sense of a distribution and disappears
when a wave packet mode is considered. It is obvious that  $\alpha^{(1)}_{\omega' \omega}$ and $\beta^{(1)}_{\omega' \omega} $ 
are finite and nonzero when $\omega \neq \omega'$, indicating different-frequency modes are also correlated. 

For $n>1$, there are no analytic expressions for $\alpha^{(n)}_{\omega' \omega}$ and $\beta^{(n)}_{\omega' \omega}$. However, for large $n$, we can
find asymptotic results:
\begin{equation}\label{fardiamonds}
\begin{split}
&\langle 0 | \hat{b}^{(0)}_{\omega} \hat{b}^{(n)}_{\omega'} | 0 \rangle \approx  \frac{1}{4 a n^2}\frac{\sqrt{\Omega \Omega'}}{\sinh(\pi \Omega)\sinh(\pi \Omega')},\\
&\langle 0 | \hat{b}^{(0)\dagger}_{\omega} \hat{b}^{(n)}_{\omega'} | 0 \rangle \approx -\frac{1}{4 a n^2}\frac{\sqrt{\Omega \Omega'}}{\sinh(\pi \Omega)\sinh(\pi \Omega')}.
\end{split}
\end{equation}
The correlation decays as $\frac{1}{n^2}$ for large $n$ but does not vanish. Contrary to the adjacent diamonds, the correlation between same-frequency modes is not dominant. 

\begin{figure}[ht!]
\centering
\includegraphics[width=9.0cm]{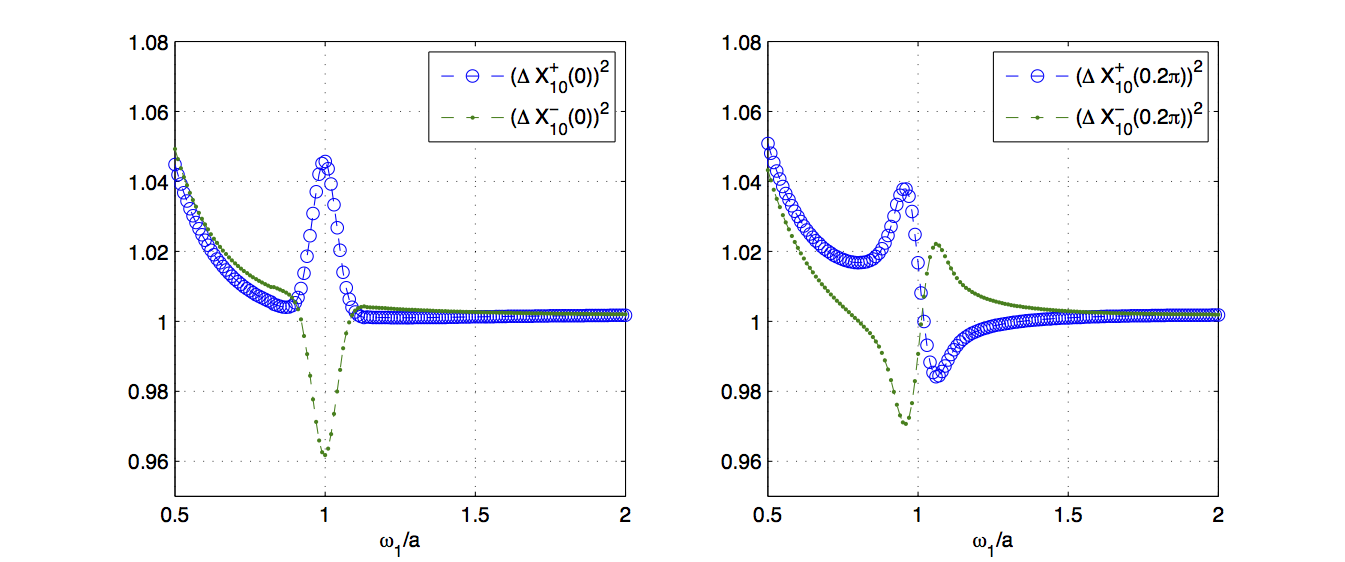}
\caption{\footnotesize (colour online). Entanglement between Gaussian modes in the first and zeroth diamond. Left: quadrature phase $\phi = 0$, right:
quadrature phase $\phi = 0.2 \pi$. The central frequency of the Gaussian mode
in the zeroth diamond is set to be $\omega_0/a = 1.0$, the bandwidth and central position are the same, $\sigma_1/a = \sigma_0/a =0.02$,
$a v_1 = a v_0 =0$. 
} 
\label{nearby}
\end{figure}

We proceed to consider localized modes instead of single-frequency modes. 
In each diamond, we construct Gaussian wave packet modes, 
\begin{equation}
\hat{b}^{(n)} = \int_0^{\infty} d \omega G(\omega;\omega_n, \sigma_n, v_n) \hat{b}^{(n)}_{\omega}, 
\end{equation}
where $G(\omega;\omega_n, \sigma_n, v_n)$ is a Gaussian wave packet
\begin{equation}
G(\omega;\omega_n, \sigma_n, v_n) = \bigg(\frac{1}{2 \pi \sigma^2_n} \bigg)^{1/4} \text{exp}\bigg\{-\frac{(\omega-\omega_n)^2}{4 \sigma^2_n}\bigg\}e^{-i \omega v_n}, \nonumber
\end{equation}
where $\omega_n$ is the central frequency, $\sigma_n$ is the bandwidth, $v_n$ is the central position of the wave packet, and we assume
$\omega_n \gg \sigma_n$. The quadratures of the Gaussian mode is defined as 
\begin{equation}
\hat{X}^{(n)}(\phi) = \hat{b}^{(n)} e^{-i \phi}+ \hat{b}^{(n)\dagger} e^{i \phi},
\end{equation}
where $\phi$ is the quadrature phase. With $\phi = 0$ and $\phi = \frac{\pi}{2}$, the quadrature $\hat{X}^{(n)}$ are analogous to the position operator and
momentum operator, respectively. Correlations between diamonds can be characterised by the variances of 
$\hat{X}_{nm}^{\pm} \equiv (\hat{X}^{(n)} \pm \hat{X}^{(m)})/\sqrt{2}$ . For example, for two-mode squeezing, 
$V(\hat{X}^{-}_{nm}(0)) < 1$ and $V(\hat{X}^{+}_{nm}(\frac{\pi}{2})) < 1$, indicating that the correlations between 
the quadratures of the two modes beat the quantum shot noise and are entangled. 

For $\phi=0$ (Fig. \ref{nearby}, left), $(\Delta X^-_{10})^2<1$ for two Gaussian modes in adjacent diamonds with the same central frequency, 
bandwidth and central position. The correlation between the two Gaussian modes beat the quantum shot noise, that is,
they are entangled. In fact, since the bandwidth is so small that the mode distributes across almost  the whole
diamond, the central position of the Gaussian mode is not so relevant. For nonzero $\phi$, e.g., $\phi=0.2 \pi$ (Fig. \ref{nearby}, right),
correlation between two Gaussian modes with different central frequency can beat the quantum shot noise. These 
properties are different from the Rindler entanglement which only exists between same-frequency modes and is 
independent of the quadrature phase. In the next
nearby diamonds, correlation between Gaussian modes with much broader bandwidth still can beat the quantum shot noise, 
although the entanglement is very small. 
That implies entanglement is stored between Gaussian modes localized in position rather than in frequency. For further away
diamonds, it is hard to see entanglement. Although Eq. (\ref{fardiamonds}) shows that the correlation has not vanished, it is very small.

In $(1+3)$-dimensional Minkowski spacetime, although we know that a mode inside a diamond is perfectly correlated with a 
corresponding mode outside the diamond, explicit expressions for these modes have not yet been found. In addition, the diamond
modes are not complete in the whole Minkowski spacetime. This can be seen by noticing that at $t=0$, the diamonds can not
cover the whole space. However, if we only consider timelike entanglement between diamonds along the $t$ axis, the method used
in this section is still valid. In realistic quantum optics experiments, a detector often detects a localized mode, e.g., a Gaussian beam with very 
narrow transverse size travelling along the $x$ axis, then the $(1+1)$-dimensional calculation provides a very good approximation 
to the $(1+3)$-dimensional case.

\section{Conclusions}\label{conclusions}

By directly calculating the Bogoliubov transformation between the diamond modes and the Minkowski modes, we show that 
the particle-number distribution in the diamond is thermal in the Minkowski vacuum. The temperature of the thermal distribution
is identical to the diamond temperature (that observed by an inertial observer at the middle of their lifetime) discovered by Martinetti and
Rovelli \cite{Rovelli03}. We interpret this temperature as the diamond temperature and show that a particular type of energy-scaled detector responds 
to the diamond temperature. The temperature is constant with respect to diamond time, but varies with respect to lab time. 
It is, therefore, clear that the diamond temperature is real and detectable. An order-of-magnitude calculation shows that 
$T_D \sim 1 K$corresponds to $\mathcal{T} \sim 10^{-11} s$ , which is challenge but potentially accessible in the lab in near future. 
We further study the timelike
entanglement between various diamonds and show that entanglement between adjacent diamonds is dominant. 

\section*{ACKNOWLEGEMENTS}
We thank S. J. Olson for useful discussions. This research was supported in part by Australian Research Council Centre of Excellence of Quantum 
Computation and Communication Technology (Project No. CE110001027). 



\end{document}